\documentclass[notitlepage,a4paper,aps,prd,onecolumn,superscriptaddress,nofootinbib,groupedaddress]{revtex4}
\usepackage{amsmath}
\usepackage{amsfonts,color}
\usepackage{amsmath}
\usepackage{amsthm}
\usepackage{pdfpages}
\usepackage{amsfonts,color}
\usepackage{amssymb,float}

\usepackage{amsmath}

\DeclareMathOperator{\arccosh}{arccosh}

\newcommand{\dd}{{\rm d}}
\usepackage[utf8]{inputenc}
\allowdisplaybreaks
\usepackage{color}
\usepackage{hyperref}
\hypersetup{
    colorlinks=true,
    linkcolor=blue,
    filecolor=magenta,      
   citecolor=blue
}

\usepackage{enumitem}

\usepackage{tikz}
\usetikzlibrary{shapes.geometric}
\usetikzlibrary{arrows.meta,arrows}

\begin{document}
\title{Pleba\'{n}ski-Demia\'{n}ski solutions with dynamical torsion and nonmetricity fields}

\author{Sebastian Bahamonde}
\email{sbahamondebeltran@gmail.com, bahamonde.s.aa@m.titech.ac.jp}
\affiliation{Department of Physics, Tokyo Institute of Technology
1-12-1 Ookayama, Meguro-ku, Tokyo 152-8551, Japan.}
\affiliation{Laboratory of Theoretical Physics, Institute of Physics, University of Tartu, W. Ostwaldi 1, 50411 Tartu, Estonia.}

\author{Jorge Gigante Valcarcel}
\email{jorge.gigante.valcarcel@ut.ee}
\affiliation{Laboratory of Theoretical Physics, Institute of Physics, University of Tartu, W. Ostwaldi 1, 50411 Tartu, Estonia.}

\author{Laur J\"arv}
\email{laur.jarv@ut.ee}
\affiliation{Laboratory of Theoretical Physics, Institute of Physics, University of Tartu, W. Ostwaldi 1, 50411 Tartu, Estonia.}

\begin{abstract}

We construct Pleba\'{n}ski-Demia\'{n}ski stationary and axisymmetric solutions with two expanding and double principal null directions in the framework of Metric-Affine gauge theory of gravity. Starting from the new improved form of the metric with vanishing cosmological constant recently achieved by Podolsk\'y and Vr\'atn\'y, we extend this form in the presence of a cosmological constant and derive the conditions under which the physical sources of the torsion and nonmetricity tensors provide dynamical contributions preserving it in Weyl-Cartan geometry. The resulting black hole configurations are characterised by the mass, orbital angular momentum, acceleration, NUT parameter, cosmological constant and electromagnetic charges of the Riemannian sector of the theory, as well as by the spin and dilation charges of the torsion and nonmetricity fields. The former is subject to a constraint representing a decoupling limit with the parameters responsible of axial symmetry, beyond which the geometry of the space-time is expected to be corrected.

\end{abstract}

\maketitle

\section{Introduction}
 
The search and analysis of exact solutions to the Einstein equations have shown to be essential in order to improve the understanding of General Relativity (GR) as the most successful and accurate theory of gravity. Among them, the well-known rotating black hole solution depending on the mass, angular momentum and electromagnetic charges turns out to be described by the Kerr-Newman geometry but it can be extended to include three additional quantities with a direct physical meaning, namely the acceleration, the NUT parameter and the cosmological constant. The resulting configuration acquires the so-called Pleba\'{n}ski-Demia\'{n}ski (PD) geometry and constitutes the most general type D solution of the Einstein-Maxwell theory with two expanding and double principal null directions \cite{Plebanski:1976gy,Stephani:2003tm,Griffiths:2005qp,Griffiths:2009dfa,Podolsky:2021zwr} (see~\cite{Kinnersley:1969zza} for the derivation in vacuum). Despite of the cumbersome mathematical structure of the solution, it possesses a significant degree of symmetry, which is realised by the existence of a nontrivial Killing-Yano tensor. In fact, even though Killing-Yano tensors do not arise straightforwardly from the isometry group, they represent a generalisation of Killing vectors and generate both the explicit and hidden symmetries of certain manifolds \cite{Frolov:2017kze,Frolov:2017whj}. In particular, the PD space-time admits a nondegenerate conformal Killing-Yano tensor that provides a separability structure for the wave and geodesic equations of massless fields, which in turn implies the complete integrability of null geodesics and the algebraically special type D symmetry of the Weyl tensor \cite{Kubiznak:2007kh,Frolov:2018pys,Frolov:2018ezx}. In this sense, the PD space-time reduces to different relevant cases when switching off any of its parameters, being the case of the Kerr-NUT-de Sitter geometry obtained when vanishing the acceleration and electromagnetic charges especially remarkable, since it turns out to describe the most general Einstein manifold with a closed nondegenerate conformal Killing-Yano tensor \cite{Houri:2007xz}. The closure condition allows then the corresponding separability structure to be extended for timelike geodesics and massive fields \cite{Frolov:2006pe,Oota:2007vx,Krtous:2018bvk}.

In the present work, we address the question under which conditions the PD solutions hold in Metric-Affine Gravity (MAG) by the presence of dynamical torsion and nonmetricity fields. Indeed, at present time there is a considerable growth of interest concerning gravitational interaction in the framework of metric-affine geometry, motivated especially by the recent progress in the search and analysis of exact solutions that enable the study of these post-Riemannian quantities at astrophysical and cosmological scales (e.g. see~\cite{Obukhov:2019fti,Bahamonde:2020fnq,Bahamonde:2021qjk,Obukhov:2020hlp,Guerrero:2020azx,Bahamonde:2021akc,Blagojevic:2020ymf,Blagojevic:2021pqp,Olmo:2019flu,Blagojevic:2017wzf,Blagojevic:2017ssv,Jimenez-Cano:2020lea,Dombriz:2021bnl,Iosifidis:2020gth,Aoki:2020zqm}). Following this line of research, we organise the article as follows. In Sec.~\ref{sec:action} we introduce metric-affine geometry and the MAG model with higher order corrections recently proposed to display black hole solutions with both dynamical torsion and nonmetricity fields. Then, in Sec.~\ref{sec:axial} we set the metric and the independent parts of the affine connection within general stationary and axisymmetric space-times. The respective spin connection turns out to admit a clear decomposition into propagating and nonpropagating modes, as well as a significantly simple form in a rotated frame, which strongly simplifies the problem and provides a systematic method to solve the field equations. This enables the study of the PD space-time in Sec.~\ref{sec:PlebDemSec}, which for the case with vanishing cosmological constant has recently been presented in terms of a new parametrisation that explicitly shows the form of all its physical parameters in the metric tensor ~\cite{Podolsky:2021zwr}. Accordingly, in Sec.~\ref{sec:4a} we consider the aforementioned parametrisation and present the full form of the PD space-time which satisfies the Einstein-Maxwell equations in the presence of a cosmological constant. The decomposition of the spin connection in terms of propagating and nonpropagating modes allows us then to obtain a new PD black hole solution with torsion and nonmetricity in Sec.~\ref{sec:4b}, under the assumption that the coupling between the spin angular momentum and the parameters responsible of axial symmetry is sufficiently small to be negligible. The main geometric effect of the physical sources of the torsion and nonmetricity fields is described in Sec.~\ref{sec:4c}. In particular, an analytical expression for the black hole and acceleration horizons as well as for the quadratic curvature invariant can be obtained by switching off the cosmological constant, which provide the specific conditions that the parameters must fulfill in order to display a regular black hole solution without any pathological inner horizon. Finally, we conclude our main results in Sec.~\ref{sec:conclusions}.

We work in the natural units $c=G=1$ with the metric signature $(+,-,-,-)$. Geometric quantities with a tilde on top denote that they are computed with respect to a general affine connection whereas quantities without a tilde denote that they are computed with respect to the Levi-Civita connection. On the other hand, Latin and Greek indices refer to anholonomic and coordinate basis, respectively. Finally, we assume Boyer-Lindquist coordinates $(t,r,\vartheta,\varphi)$.

\section{Metric-Affine Gravity with dynamical torsion and nonmetricity}\label{sec:action}

In the realm of GR, the description of gravity and its interaction with matter fields is successfully achieved by establishing a geometric correspondence between the curvature tensor of a Riemannian manifold and the energy-momentum tensor of matter that respects the invariance under diffeomorphisms \cite{Wald:1984rg}. The measurement of distances and the parallel transport of tensors are then completely determined by the metric tensor, in virtue of the Fundamental Theorem of Riemannian geometry that sets the corresponding affine connection as the Levi-Civita connection
\begin{equation}
\Gamma^{\lambda}\,_{\mu \nu}=\frac{1}{2}\,g^{\lambda \rho}\left(\partial_{\mu}g_{\nu \rho}+\partial_{\nu}g_{\mu \rho}-\partial_{\rho}g_{\mu \nu}\right)\,.
\end{equation}

Nevertherless, the absence of finite spinor representations for the diffeomorphism group requires a principal bundle connection $\omega_{\mu} \in \mathfrak{so}(1,3)$ to introduce the dynamics of spinor fields into this geometrical scheme, which leads to a gauge characterisation of the geometry of the space-time \cite{Hehl:1976kj,ponomarev2017gauge}. In this sense, the application of the gauge procedure to the external degrees of freedom consisting of translations, rotations, dilations and shears by means of the affine group $A(4,R) = R^{4} \rtimes GL(4,R)$ turns out to provide a fundamental relation between the gauge curvatures
\begin{eqnarray}
G_{ab\mu}&=&\partial_{\mu}g_{ab}-g_{ac}\,\omega^{c}\,_{b\mu}-g_{bc}\,\omega^{c}\,_{a\mu}\,,
\\
F^{a}\,_{\mu\nu}&=&\partial_{\mu}e^{a}\,_{\nu}-\partial_{\nu}e^{a}
\,_{\mu}+\omega^{a}\,_{b\mu}\,e^{b}\,_{\nu}-\omega^{a}\,_{b\nu}\,e^{b}\,_{\mu}\,,
\\
F^{a}\,_{b\mu\nu}&=&\partial_{\mu}\omega^{a}\,_{b\nu}
-\partial_{\nu}\omega^{a}\,_{b\mu}+\omega^{a}\,_{c\mu}
\,\omega^{c}\,_{b}\,_{\nu}-\omega^{a}\,_{c\nu}
\,\omega^{c}\,_{b\mu}\,,
\end{eqnarray}
and the nonmetricity, torsion and curvature tensors of an affinely connected metric manifold \cite{Hehl:1994ue,Blagojevic:2013xpa,Cabral:2020fax}:
\begin{eqnarray}
G_{ab\mu}&=&\,g_{a c}g_{b d}e^{c\lambda}e^{d\rho}Q_{\mu\lambda\rho},
\\
F^{a}\,_{\mu\nu}&=&e^{a}\,_{\lambda}T^{\lambda}\,_{\nu\mu}\,,
\\
F^{a}\,_{b\mu\nu}&=&g_{b c}\,e^{a}\,_{\lambda}e^{c\rho}\tilde{R}^{\lambda}\,_{\rho\mu\nu}\,,
\end{eqnarray}
where the coframe field, the spin connection and the local metric satisfy the relations\footnote{Note that $e_{a}\,^{\mu}$ denotes the inverse of the coframe, which satisfies the orthonormality properties $e^{a}\,_{\mu}e_{b}\,^{\mu}=\delta^{a}\,_{b}$ and $e^{a}\,_{\mu}e_{a}\,^{\nu}=\delta_{\mu}\,^{\nu}$.}:
\begin{eqnarray}
g_{\mu \nu}&=&e^{a}\,_{\mu}\,e^{b}\,_{\nu}\,g_{a b}\,,\label{vierbein_def}\\
\omega^{a}\,_{b\mu}&=&e^{a}\,_{\lambda}\,e_{b}\,^{\rho}\,\tilde{\Gamma}^{\lambda}\,_{\rho \mu}+e^{a}\,_{\lambda}\,\partial_{\mu}\,e_{b}\,^{\lambda}\,.\label{anholonomic_connection}
\end{eqnarray}

The physical implications of such a construction are highly relevant since the main conclusion is that a gauge approach to gravity naturally displays the nonmetricity, torsion and curvature tensors as the gauge curvatures of the model carrying the gravitational interaction. Accordingly, a gauge invariant Lagrangian can be constructed from these tensors to introduce the extended properties of the gravitational field, in such a way that the propagating behaviour of torsion and nonmetricity requires at least the presence of quadratic order corrections in the curvature tensor. This feature provides in general a rich gravitational spectrum and substantially increases the complexity of the theory, since the algebraic structure of the external affine group allows the definition of a large number of gauge invariants, in contrast with the regular Yang-Mills theories of internal compact symmetry groups \cite{Yang:1954ek}. In particular, the most general parity conserving quadratic Lagrangian includes 11+3+4 irreducible modes from curvature, torsion and nonmetricity, respectively \cite{McCrea:1992wa}. A strong but still meaningful simplification is achieved when the nonmetricity tensor is fully ascribed to its trace irreducible part, in such a way that the resulting affine connection includes the torsion tensor and the so-called Weyl vector:\begin{align}
    T^{\mu}\,_{\nu\rho}&=2\tilde{\Gamma}^{\mu}\,_{[\nu\rho]}\,,\quad
    Q_{\mu\nu\rho}=\tilde{\nabla}_{\mu}g_{\nu\rho}\,,\quad
    W_{\mu}=\frac{1}{4}\,Q_{\mu\nu}\,^{\nu}\,,
\end{align}
being the resulting geometry known as Weyl-Cartan geometry \cite{Blagojevic:2002du}. In this context, the main geometric implications provided by the torsion and the nonmetricity tensors within the manifold are the breaking of standard parallelograms and the change of the lengths of vectors along
infinitesimal paths, respectively. The affine connection is then decomposed into a Riemannian Levi-Civita part determined by the metric tensor and a distortion tensor
\begin{equation}\label{affineconnection}
\tilde{\Gamma}^{\lambda}\,_{\mu \nu}=\Gamma^{\lambda}\,_{\mu \nu}+N^{\lambda}\,_{\mu \nu}\,,
\end{equation}
which encodes the torsion and nonmetricity tensors as the sum of the so-called contortion and disformation tensors
\begin{equation}
N^{\lambda}\,_{\rho\mu}=K^{\lambda}\,_{\rho\mu}+L^{\lambda}\,_{\rho\mu}\,,
\end{equation}
with
\begin{eqnarray}
K^{\lambda}\,_{\mu \nu}&=&\frac{1}{2}\left(T^{\lambda}\,_{\mu \nu}-T_{\mu}\,^{\lambda}\,_{\nu}-T_{\nu}\,^{\lambda}\,_{\mu}\right)\,,\\
L^{\lambda}\,_{\mu \nu}&=&\frac{1}{2}\left(Q^{\lambda}\,_{\mu \nu}-Q_{\mu}\,^{\lambda}\,_{\nu}-Q_{\nu}\,^{\lambda}\,_{\mu}\right)\,.
\end{eqnarray}

Likewise, the change on the commutation relations of the covariant derivatives
\begin{equation}
[\tilde{\nabla}_{\mu},\tilde{\nabla}_{\nu}]\,v^{\lambda}=\tilde{R}^{\lambda}\,_{\rho \mu \nu}\,v^{\rho}+T^{\rho}\,_{\mu \nu}\,\tilde{\nabla}_{\rho}v^{\lambda}\,,
\end{equation}
introduces the mentioned post-Riemannian corrections in the curvature tensor
\begin{equation}\label{totalcurvature}
\tilde{R}^{\mu}\,_{\nu\rho\sigma}=\partial_{\rho}\tilde{\Gamma}^{\mu}\,_{\nu\sigma}-\partial_{\sigma}\tilde{\Gamma}^{\mu}\,_{\nu\rho}+\tilde{\Gamma}^{\mu}\,_{\lambda\rho}\tilde{\Gamma}^{\lambda}\,_{\nu\sigma}-\tilde{\Gamma}^{\mu}\,_{\lambda\sigma}\tilde{\Gamma}^{\lambda}\,_{\nu\rho}\,,
\end{equation}
which generalises the Bianchi identities of Riemannian geometry in the presence of torsion and nonmetricity
\begin{eqnarray}\tilde{R}^{\lambda}\,_{[\mu \nu \rho]}&=&\tilde{\nabla}_{[\mu}T^{\lambda}\,_{\rho\nu]}+T^{\sigma}\,_{[\mu\rho}\,T^{\lambda}\,_{\nu] \sigma}\,,\\\label{curvbianchi1}
\tilde{\nabla}_{[\sigma |}\tilde{R}^{\lambda}\,_{\rho | \mu \nu]}&=&T^{\omega}\,_{[\sigma \mu |}\tilde{R}^{\lambda}\,_{\rho \omega | \nu]}\,,\\\label{curvbianchi2}
\tilde{R}^{\left(\lambda\rho\right)}\,_{\mu\nu}&=&\tilde{\nabla}_{[\nu}Q_{\mu]}\,^{\lambda\rho}+\frac{1}{2}\,T^{\sigma}\,_{\mu\nu}Q_{\sigma}\,^{\lambda\rho}\,,\label{nonmetricitybianchi}
\end{eqnarray}
and implies the fulfillment of the following relation for the curvature tensor under the exchange of the pair of indices:
\begin{equation}\tilde{R}_{[\lambda \rho] \mu \nu}-\tilde{R}_{[\mu \nu] \lambda \rho}=\frac{3}{2}\left(\tilde{R}_{\lambda [\rho \mu \nu]}+\tilde{R}_{\rho [\mu \lambda \nu]}+\tilde{R}_{\mu [\rho \lambda \nu]}+\tilde{R}_{\nu [\rho \mu \lambda]}\right)+\left(\tilde{R}_{(\lambda \mu) \rho \nu}+\tilde{R}_{(\lambda \nu) \mu \rho}+\tilde{R}_{(\rho \mu) \nu \lambda}+\tilde{R}_{(\rho \nu) \lambda \mu}\right)\,.\end{equation}

Additionally, the existence of a symmetric component (\ref{nonmetricitybianchi}) of the curvature tensor allows the definition of two independent traces constructed from the distinct pair of indices, the Ricci and co-Ricci tensors
\begin{eqnarray}\label{Riccitensor}
\tilde{R}_{\mu\nu}&=&\tilde{R}^{\lambda}\,_{\mu \lambda \nu}\,,\\
\label{co-Riccitensor}
\hat{R}_{\mu\nu}&=&\tilde{R}_{\mu}\,^{\lambda}\,_{\nu\lambda}\,,
\end{eqnarray}
and a third one from the first pair of indices that encodes the change of lengths under parallel transport, namely the homothetic curvature tensor
\begin{equation}\label{homothetic}
\tilde{R}^{\lambda}\,_{\lambda\mu\nu}=4\nabla_{[\nu}W_{\mu]}\,.
\end{equation}

The trace of the first Bianchi identity~(\ref{curvbianchi1}) expresses then the antisymmetric part of the Ricci tensor in terms of the homothetic component provided by the Weyl vector and the respective corrections depending on the torsion tensor
\begin{equation}\tilde{R}_{[\mu \nu]}=\frac{1}{2}\tilde{R}^\lambda\,_{\lambda\mu\nu}+\tilde{\nabla}_{[\mu}T^{\lambda}\,_{\nu]\lambda}+\frac{1}{2}\tilde{\nabla}_{\lambda}T^{\lambda}\,_{\mu\nu}-\frac{1}{2}T^{\lambda}\,_{\rho\lambda}T^{\rho}\,_{\mu\nu}\,.\end{equation}

Following these lines, the formulation of models that allow a phenomenological assessment of MAG by the search and analysis of exact solutions turns out to be especially desirable. In this regard, it has been recently shown that the following model defined on Weyl-Cartan geometry displays black hole configurations with both dynamical torsion and nonmetricity fields, which are also compatible with the parity violating sector of the theory \cite{Bahamonde:2020fnq,Obukhov:2020hlp} (see also~\cite{Blagojevic:2021pqp} for a consistent thermodynamic analysis in the absence of nonmetricity):
\begin{eqnarray}\label{LagrangianIrreducible}
S &=& \frac{1}{64\pi}\int \dd^4x \sqrt{-\,g}
\left.\Bigl[2\Lambda
-\,4R-6d_{1}\tilde{R}_{\lambda\left[\rho\mu\nu\right]}\tilde{R}^{\lambda\left[\rho\mu\nu\right]}-9d_{1}\tilde{R}_{\lambda\left[\rho\mu\nu\right]}\tilde{R}^{\mu\left[\lambda\nu\rho\right]}+8\,d_{1}\tilde{R}_{\left[\mu\nu\right]}\tilde{R}^{\left[\mu\nu\right]}
\Bigr.
\right.
\nonumber\\
& &
\left.
\Bigl.
\;\;\;\;\;\;\;\;\;\;\;\;\;\;\;\;\;\;\;\;\;\;\;\;\;\;\;\;\;+\,\frac{1}{8}\left(32e_{1}+13d_{1}\right)\tilde{R}^{\lambda}\,_{\lambda\mu\nu}\tilde{R}^{\rho}\,_{\rho}\,^{\mu\nu}-7d_{1}\tilde{R}_{\left[\mu\nu\right]}\tilde{R}^{\lambda}\,_{\lambda}\,^{\mu\nu}\Bigr]\right.\,,
\end{eqnarray}
in such a way that the affine group $A(4,R)$ reduces to the inhomogeneous Weyl group $IW(1,3)$, characterised by a gauge connection
\begin{equation}
    A_{\mu}=e^{a}\,_{\mu}P_{a}+\omega^{[a b]}\,_{\mu}J_{a b}+\frac{1}{4}\,\omega^{(a b)}_{\mu}\eta_{a b}D\,,
\end{equation}
where $P_{a}$, $J_{a b}$ and $D$ are respectively the generators of translations, Lorentz rotations and dilations, whereas \;\, $\eta^{ab}=\textrm{diag}\left(1,-\,1,-\,1,-\,1\right)$. Therefore, the generators of the model satisfy the commutation relations
\begin{eqnarray}
\left[P_{a},P_{b}\right]&=&\left[D,J_{ab}\right]=\left[D,D\right]=0\,,\\
\left[D,P_{a}\right]&=&i\,P_{a}\,,\\
\left[P_{a},J_{bc}\right]&=&i\,\eta_{a[b}\,P_{c]}\,,\\
\left[J_{ab},J_{cd}\right]&=&\frac{i}{2}\,\left(\eta_{ad}\,J_{bc}+\eta_{cb}\,J_{ad}-\eta_{db}\,J_{ac}-\eta_{ac}\,J_{bd}\right)\,,
\end{eqnarray}
and define a Lie algebra of the group $IW(1,3)$ in any of its induced representations, e.g. in the spinor representation:
\begin{equation}
J_{ab}=\frac{i}{8}\left[\gamma_a,\gamma_b\right]\,, \quad P_{a}=\frac{i}{2}\left(1+\gamma^{5}\right)\gamma_{a}\,, \quad D=\frac{i}{2}\gamma^{5}\,.
\end{equation}

As can be seen from Expression~(\ref{LagrangianIrreducible}), all the geometric corrections lie on the torsion and nonmetricity fields, in such a way that Einstein's theory in the presence of a cosmological constant $\Lambda$ is fully recovered when they vanish. This property is appropriately expressed in the field equations, which generalise the vacuum Einstein equations in the presence of a cosmological constant with dynamical contributions of torsion and nonmetricity as follows:
\begin{eqnarray}\label{field_eq1}
0 &=& \,G_{\mu}\,^{\nu}+\Lambda \delta_{\mu}\,^{\nu}+8\pi\tilde{\mathcal{L}}\,\delta_{\mu}\,^{\nu}+d_{1}\left(\tilde{R}^{\nu}\,_{\lambda\rho\mu}\tilde{R}^{\left[\lambda\rho\right]}+\tilde{R}_{\lambda}\,^{\nu}\,_{\mu\rho}\hat{R}^{\left[\lambda\rho\right]}+\tilde{R}_{\lambda\mu}\tilde{R}^{\left[\nu\lambda\right]}+\hat{R}_{\lambda\mu}\hat{R}^{\left[\nu\lambda\right]}\right)
\nonumber\\
&&+\frac{d_1}{4}\tilde{R}_{\lambda\rho\sigma\mu}
\Bigl[
\left(4\tilde{R}^{\left[\nu\sigma\right]\lambda\rho}-2\tilde{R}^{\left[\lambda\rho\right]\nu\sigma}-\tilde{R}^{\left[\lambda\nu\right]\rho\sigma}-\tilde{R}^{\left[\rho\sigma\right]\lambda\nu}-\tilde{R}^{\left[\rho\nu\right]\sigma\lambda}-\tilde{R}^{\left[\sigma\lambda\right]\rho\nu}\right)
\Bigr.\,
\nonumber\\
\Bigl.
&&-\,4\left(\tilde{R}^{\left(\rho\nu\right)\lambda\sigma}-\tilde{R}^{\left(\lambda\sigma\right)\rho\nu}+\tilde{R}^{\left(\lambda\nu\right)\rho\sigma}-\tilde{R}^{\left(\rho\sigma\right)\lambda\nu}\right)
\Bigr]-2e_{1}\tilde{R}^{\lambda}\,_{\lambda\sigma\mu}\tilde{R}^{\rho}\,_{\rho}\,^{\sigma\nu}\,,
\end{eqnarray}
\begin{eqnarray}\label{field_eq2}
0&=&2d_{1}
\Bigl[
\nabla_{\rho}\left(g^{\mu\nu}\tilde{R}^{\left[\lambda\rho\right]}-g^{\lambda\nu}\hat{R}^{\left[\mu\rho\right]}+g^{\lambda\rho}\hat{R}^{\left[\mu\nu\right]}-g^{\mu\rho}\tilde{R}^{\left[\lambda\nu\right]}\right)+N^{\rho\mu}\,_{\rho}\tilde{R}^{\left[\lambda\nu\right]}-N^{\rho\lambda}\,_{\rho}\hat{R}^{\left[\mu\nu\right]}
\Bigr.
\nonumber\\
\Bigl.
&&+N^{\nu\lambda}\,_{\rho}\hat{R}^{\left[\mu\rho\right]}-N^{\nu\mu}\,_{\rho}\tilde{R}^{\left[\lambda\rho\right]}+N^{\mu}\,_{\rho}\,^{\lambda}\hat{R}^{\left[\rho\nu\right]}-N^{\lambda}\,_{\rho}\,^{\mu}\tilde{R}^{\left[\rho\nu\right]}
\Bigr]
\nonumber\\
&&+\frac{d_1}{2}\Bigl(\nabla_{\rho}+W_{\rho}\Bigr)
\Bigl[
\left(4\tilde{R}^{\left[\rho\nu\right]\lambda\mu}-2\tilde{R}^{\left[\lambda\mu\right]\rho\nu}-\tilde{R}^{\left[\mu\nu\right]\lambda\rho}+\tilde{R}^{\left[\lambda\nu\right]\mu\rho}-\tilde{R}^{\left[\lambda\rho\right]\mu\nu}+\tilde{R}^{\left[\mu\rho\right]\lambda\nu}\right)
\Bigr.
\nonumber\\
\Bigl.
&&-\,4\left(\tilde{R}^{\left(\mu\nu\right)\lambda\rho}-\tilde{R}^{\left(\lambda\rho\right)\mu\nu}+\tilde{R}^{\left(\nu\lambda\right)\mu\rho}-\tilde{R}^{\left(\rho\mu\right)\lambda\nu}\right)
\Bigr]
-4e_{1}g^{\lambda \mu}\nabla_{\rho}\tilde{R}_{\sigma}\,^{\sigma\rho\nu}
\nonumber\\
&&+\frac{d_1}{2}N^{\lambda}\,_{\sigma\rho}
\Bigl[
\left(4\tilde{R}^{\left[\rho\nu\right]\sigma\mu}-2\tilde{R}^{\left[\sigma\mu\right]\rho\nu}-\tilde{R}^{\left[\mu\nu\right]\sigma\rho}+\tilde{R}^{\left[\sigma\nu\right]\mu\rho}-\tilde{R}^{\left[\sigma\rho\right]\mu\nu}+\tilde{R}^{\left[\mu\rho\right]\sigma\nu}\right)
\Bigr.
\nonumber\\
\Bigl.
&&-\,4\left(\tilde{R}^{\left(\mu\nu\right)\sigma\rho}-\tilde{R}^{\left(\sigma\rho\right)\mu\nu}+\tilde{R}^{\left(\nu\sigma\right)\mu\rho}-\tilde{R}^{\left(\rho\mu\right)\sigma\nu}\right)+4g^{\mu\nu}\tilde{R}^{\left[\sigma\rho\right]}
\Bigr]
\nonumber\\
&&+\frac{d_1}{2}N^{\mu}\,_{\sigma\rho}
\Bigl[
\left(4\tilde{R}^{\left[\rho\nu\right]\lambda\sigma}-2\tilde{R}^{\left[\lambda\sigma\right]\rho\nu}-\tilde{R}^{\left[\sigma\nu\right]\lambda\rho}+\tilde{R}^{\left[\lambda\nu\right]\sigma\rho}-\tilde{R}^{\left[\lambda\rho\right]\sigma\nu}+\tilde{R}^{\left[\sigma\rho\right]\lambda\nu}\right)
\Bigr.
\nonumber\\
\Bigl.
&&-\,4\left(\tilde{R}^{\left(\sigma\nu\right)\lambda\rho}-\tilde{R}^{\left(\lambda\rho\right)\sigma\nu}+\tilde{R}^{\left(\nu\lambda\right)\sigma\rho}-\tilde{R}^{\left(\sigma\rho\right)\lambda\nu}\right)-4g^{\lambda\nu}\hat{R}^{\left[\sigma\rho\right]}
\Bigr]\,,
\end{eqnarray}
where $\tilde{\mathcal{L}}$ represents the quadratic order of the Lagrangian density. The highly nonlinear character and the large number of degrees of freedom contained in the field equations require additional assumptions to provide a solution. In this regard, it is been recently shown that there are consistent solutions to the field equations of MAG where the spin connection in a rotated frame $\vartheta^{a}=\Lambda^{a}\,_{b}\,e^{b}$ can be strongly simplified and decomposed as a Minkowski part that does not depend on any of the physical parameters of the solution plus the corresponding dynamical contributions of the torsion and nonmetricity fields \cite{Bahamonde:2020fnq,Bahamonde:2021qjk} (see \cite{Cembranos:2016gdt,Cembranos:2017pcs} for the respective structure of the solutions in the Poincar\'e gauge framework of MAG). Then, the spin connection denoted by hat in the rotated frame can be expressed in terms of the torsion and nonmetricity tensors, according to the correspondence~(\ref{anholonomic_connection}) with the affine connection. The former is written with a bar or circle on top, to indicate its dynamical or nondynamical contribution to the field strength tensors, respectively
\begin{equation}\label{SCtotal}
    \hat{\omega}^{a b}\,_{\mu}=\sum_{i=1}^{3}\hat{\omega}_{i}^{a b}\,_{\mu}\,,
\end{equation}
where
\begin{align}
    \hat{\omega}_{1}^{a b}\,_{\mu}&=\vartheta^{a}\,_{\lambda}\vartheta^{b \rho}\left(\Gamma^{\lambda}\,_{\rho \mu}+\mathring{K}_{1}^{\lambda}\,_{\rho\mu}\right)+\vartheta^{a}\,_{\lambda}\partial_{\mu}\vartheta^{b \lambda}\,,\label{SCpart1}\\
    \hat{\omega}_{2}^{a b}\,_{\mu}&=\vartheta^{a}\,_{\lambda}\vartheta^{b \rho}\left(\mathring{K}_{2}^{\lambda}\,_{\rho\mu}+L^{\lambda}\,_{\rho\mu}\right)\,,\label{SCpart2}\\
    \hat{\omega}_{3}^{a b}\,_{\mu}&=\vartheta^{a}\,_{\lambda}\vartheta^{b \rho}\bar{K}^{\lambda}\,_{\rho\mu}\label{SCpart3}\,.
\end{align}

Thereby, the problem is reduced to the search of the three main pieces of the spin connection. Specifically, the first component $\hat{\omega}_{1}$ is identified as the Minkowski part, which does not give rise to any dynamical contribution to the field strength tensors and represents the correspondence of the gravitational action with GR:
\begin{eqnarray}\label{MinkowskiSP}
\hat{\omega}_{1}=
\frac{1}{2}\bigl(J_{\hat{1}\hat{2}}-J_{\hat{0}\hat{2}}\bigr)\,\dd\vartheta +\frac{1}{2}\sin\vartheta \left[\bigl(J_{\hat{1}\hat{3}}-J_{\hat{0}\hat{3}}\bigr)+2\cot\vartheta J_{\hat{2}\hat{3}}\right]\dd\varphi\,,
\end{eqnarray}
On the other hand, the components $\hat{\omega}_{2}$ and $\hat{\omega}_{3}$ describe the dynamical contributions of the nonmetricity and torsion tensors, respectively. Such a decomposition is also reflected in the field equations, since they identically vanish if the spin connection is reduced to the first part associated with a vanishing contribution to the field strength tensor
\begin{equation}\label{vanishingantisymcurv}
\vartheta^{b}\,_{[\rho|}\mathcal{F}_{1}^{a}\,_{b|\mu\nu]}=0\,,
\end{equation}
denoted by a calligraphic letter to indicate that it is expressed in the rotated frame, whereas the symmetric and antisymmetric components of the Expression (\ref{field_eq2}) involving the indices $\lambda$ and $\mu$ turn out to provide the remaining parts related to the dynamical nonmetricity and torsion tensors, respectively.

The evaluation of the field equations for the mentioned decomposition can be realised in general axisymmetric space-times and, more specifically, in the PD geometry, as shown below.

\section{Axial symmetry in Metric-Affine Gravity}\label{sec:axial}

The explicit symmetries of stationary and axisymmetric space-times are generated by two Killing vectors $\partial_{t}$ and $\partial_{\varphi}$, which preserve the local metric structure under time translations and rotations around a symmetry axis, respectively. In particular, the latter acts as the generator of the rotation group SO(2) and defines a regular two-dimensional timelike surface of fixed points where it vanishes \cite{Stephani:2003tm,Ortin:2015hya}. The lack of orthogonality between these Killing vectors induces a nonvanishing off-diagonal component $g_{t\varphi}$ in the metric tensor, in such a way that it is possible to write the most general stationary and axisymmetric space-time up to a gauge choice as follows \cite{hartle1967variational}:
\begin{align}\label{eq:metric}
    \dd s^2&=\Psi_1(r,\vartheta)\, \dd t^2-\frac{\dd r^2}{\Psi_2(r,\vartheta)}- r^2 \Psi_3(r,\vartheta)\,\dd\vartheta^2- \Psi_3(r,\vartheta) \Psi_{5}^{2}(r,\vartheta)  r^2 \sin ^2\vartheta \, \dd\varphi^2\nonumber\\
    &\,\,\,\,+2  r^2\,\Psi_3(r,\vartheta) \Psi_4(r,\vartheta)\Psi_5(r,\vartheta)\sin ^2\vartheta\, \dd t \, \dd\varphi\,.
                            \end{align}
                            
Accounting for the nonorthogonality property of the Killing vectors in the respective off-diagonal terms $e^{\hat{0}}\,_{\varphi}$ and $e^{\hat{3}}\,_{t}$ and assuming that the rest of the off-diagonal components are equal to zero, the tetrad field reproducing the above metric according to~(\ref{vierbein_def}) turns out to contain an extra function related to a remaining gauge freedom under Lorentz boosts
\begin{align}
      e^a{}_\mu &=\left(
\begin{array}{cccc}
 \sqrt{\Psi_1(r,\vartheta)+\frac{K_1^{2}(r,\vartheta)}{r^2 \Psi_3(r,\vartheta)}} & 0 & 0 & r\Psi_3(r,\vartheta)\Psi_5(r,\vartheta)K_2(r,\vartheta)\sin\vartheta \\
 0 & \frac{1}{\sqrt{\Psi_2(r,\vartheta)}} & 0 & 0 \\
 0 & 0 & r \sqrt{\Psi_3(r,\vartheta)} & 0 \\
 \frac{K_1(r,\vartheta)}{r \sqrt{\Psi_3(r,\vartheta)}} & 0 & 0 & r\Psi_5(r,\vartheta)\sin\vartheta\sqrt{\Psi_3(r,\vartheta) \left(1+\Psi_3(r,\vartheta)K_2^{2}(r,\vartheta)\right)} \\
\end{array}
\right)\,,\label{tetrad1}
\end{align}
where the functions $K_1$ and $K_2$ are constrained according to the expression
\begin{equation}\label{K5}
K_1(r,\vartheta) =r \Psi_3(r,\vartheta) \left(K_2(r,\vartheta)\sqrt{\Psi_{1}(r,\vartheta)+r^2\Psi_{3}(r,\vartheta)\Psi_{4}^{2}(r,\vartheta)\sin^2\vartheta}-r\Psi_{4}(r,\vartheta)\sin\vartheta\sqrt{1+\Psi_{3}(r,\vartheta)K_{2}^{2}(r,\vartheta)}\,\right)\,.
\end{equation}

An independent stationary and axially symmetric spin connection being invariant under the action of the two Killing vectors acquires the general form
\begin{equation}
    \omega_{\mu}=\omega_{\mu}(r,\vartheta)\,,
\end{equation}
which means that in the present case of Weyl-Cartan geometry the forms of the torsion and nonmetricity tensors are described by twenty four and four arbitrary components depending on the radial and polar coordinates, respectively:
\begin{eqnarray}
    T_{\lambda\mu\nu}&=&T_{\lambda\mu\nu}(r,\vartheta)\,,\\ W_{\mu}&=&W_{\mu}(r,\vartheta)\,.
\end{eqnarray}

In this sense, it is clear that the application of the explicit isometries on the spin connection does not restrict its form to any special subclass of stationary and axisymmetric connections. In particular, a Minkowski spin connection (\ref{MinkowskiSP}) must additionally fulfill Eq.~(\ref{vanishingantisymcurv}) in a certain rotated frame, in order to preserve its Minkowski values in a general stationary and axisymmetric space-time. The search of the gauge transformation preserving this special form can then be addressed by a set of Lorentz boosts and space rotations, parametrised as
\small{
\begin{align}\label{eq:lambda}
    \Lambda^a{}_b&=\left(
\begin{array}{cccc}
 1 & 0 & 0 & 0 \\
 0 & \cos\alpha_3 & -\sin\alpha_3 & 0 \\
 0 & \sin\alpha_3 & \cos\alpha_3 & 0 \\
 0 & 0 & 0 & 1 \\
\end{array}
\right)\times\left(
\begin{array}{cccc}
 1 & 0 & 0 & 0 \\
 0 & \cos\alpha_2 & 0 & -\sin\alpha_2 \\
 0 & 0 & 1 & 0 \\
 0 & \sin\alpha_2 & 0 & \cos\alpha_2 \\
\end{array}
\right)\times\left(
\begin{array}{cccc}
 1 & 0 & 0 & 0 \\
 0 & 1 & 0 & 0 \\
 0 & 0 & \cos\alpha_1 & -\sin\alpha_1 \\
 0 & 0 & \sin\alpha_1 & \cos\alpha_1 \\
\end{array}
\right)\nonumber\\
&\,\,\,\,\,\,\,\times \left(
\begin{array}{cccc}
 \cosh\beta_3 & 0 & 0 & \sinh\beta_3 \\
 0 & 1 & 0 & 0 \\
 0 & 0 & 1 & 0 \\
 \sinh\beta_3 & 0 & 0 & \cosh\beta_3 \\
\end{array}
\right)\times\left(
\begin{array}{cccc}
 \cosh\beta_2 & 0 & \sinh\beta_2 & 0 \\
 0 & 1 & 0 & 0 \\
 \sinh\beta_2 & 0 & \cosh\beta_2 & 0 \\
 0 & 0 & 0 & 1 \\
\end{array}
\right)\times\left(
\begin{array}{cccc}
 \cosh\beta_1 & \sinh\beta_1 & 0 & 0 \\
 \sinh\beta_1 & \cosh\beta_1 & 0 & 0 \\
 0 & 0 & 1 & 0 \\
 0 & 0 & 0 & 1 \\
\end{array}
\right)\,.
\end{align}}\normalsize
Furthermore, this matrix can be split into two different transformations that preserve the Minkowski form of the spin connection under the explicit spherical and axial symmetries, namely
\begin{equation}
\Lambda^{a}\,_{b}=\Lambda^{a}\,_{c}\big|_{\rm axi}\Lambda^{c}\,_{b}\big|_{\rm sph}\,.
\end{equation}
Specifically, the Lorentz boost provided by the parameter
\begin{equation}\label{tildebeta1}
    \beta_{1}=\arccosh\left[\frac{\Psi_{1}^{1/2}(r,\vartheta)\,\Psi_{2}^{1/2}(r,\vartheta)+1}{2\,\Psi_{1}^{1/4}(r,\vartheta)\,\Psi_{2}^{1/4}(r,\vartheta)}\right]\,,
\end{equation}
is compatible with such a form and respects the regularity conditions of the torsion tensor in the spherically symmetric case obtained when $K_{2}(r,\vartheta)=\Psi_{4}(r,\vartheta)=0$ and $\Psi_{3}(r,\vartheta)=\Psi_{5}(r,\vartheta)=1$ \cite{Bahamonde:2020fnq}. The transition to axially symmetric configurations requires then to solve Eq.~(\ref{vanishingantisymcurv}) for a general set of values of the metric and tetrad functions, which can be totally accomplished with two additional angles $\alpha_2$ and $\beta_3$ besides the previous value for the parameter $\beta_1$:
 \begin{align}\label{eq1}
 0&=\cos \alpha_2 \sinh \beta_1 \sinh \beta_3+\sin \alpha_2 \cosh \beta_1\,,\\
    0&=\sqrt{r^2\Psi_1(r,\vartheta)\Psi_3(r,\vartheta)+K_{1}^{2}(r,\vartheta)}(\cos \alpha_2 \cosh \beta_1 \sinh \beta_3+\sin \alpha_2 \sinh \beta_1)+K_1(r,\vartheta)\cos\alpha_2 \cosh \beta_3\,.\label{eq2}
\end{align}
Then, the system of equations~\eqref{eq1}-\eqref{eq2} is solved by setting
\begin{eqnarray}\label{angle1}
\alpha_2&=&\arccos\left[\sqrt{1-\frac{K_{1}^{2}(r,\vartheta)\sinh^2\beta_1}{r^2 \Psi_1(r,\vartheta) \Psi_3(r,\vartheta)}}\,\right]\,,\\
\beta_3&=&\arccosh\left[\sqrt{\frac{r^2 \Psi_1(r,\vartheta) \Psi_3(r,\vartheta)+K_{1}^{2}(r,\vartheta)}{r^2 \Psi_1(r,\vartheta) \Psi_3(r,\vartheta)-K_{1}^{2}(r,\vartheta)\sinh^2\beta_1}}\,\right]\,,\label{angle2}
\end{eqnarray}
which gives rise to the following Lorentz matrix:
\begin{align}\label{LorentzMatrixTransformation}
    \Lambda^a{}_b&=\left(
    \begin{array}{cccc}
    \cosh\beta_{3} & 0 & 0 & \sinh\beta_{3}\\
    -\sin\alpha_{2}\sinh\beta_{3} & \cos\alpha_{2} & 0 & -\sin\alpha_{2}\cosh\beta_{3}\\
    0 & 0 & 1 & 0\\
    \cos\alpha_{2}\sinh\beta_{3} & \sin\alpha_{2} & 0 & \cos\alpha_{2}\cosh\beta_{3}\\
    \end{array}
    \right)\times  \left(
    \begin{array}{cccc}
    \cosh\beta_1 & \sinh\beta_1 & 0 & 0\\
    \sinh\beta_1 & \cosh\beta_1 & 0 & 0\\
    0 & 0 & 1 & 0\\
    0 & 0 & 0 & 1\\
    \end{array}
    \right)\,.
\end{align}
Clearly, if $K_1(r,\vartheta)=0$ the two angles $\alpha_2$ and $\beta_3$ vanish, meaning that a tetrad with an off-diagonal term $e^{\hat{3}}\,_{t}=0$ does not require any additional local Lorentz transformation to solve~Eq.(\ref{vanishingantisymcurv}) in a general axially symmetric space-time. In this sense, the choice of a specific value for this function is just a matter of the gauge freedom contained in the tetrad field.

\section{Pleba\'{n}ski-Demia\'{n}ski space-time}\label{sec:PlebDemSec}

This section shall be devoted to the search of PD solutions within the MAG model described by the Expression~(\ref{LagrangianIrreducible}). First, in Sec.~\ref{sec:4a} we will present a new form of the PD metric by extending the result of~\cite{Podolsky:2021zwr} in the presence of a cosmological constant. Then, in Sec.~\ref{sec:4b} we shall assume the decomposition~(\ref{SCpart1})-(\ref{SCpart3}) of the post-Riemannian spin connection to solve the respective field equations with dynamical torsion and nonmetricity fields. Finally, from the final form of the solution, the main differences provided by the physical sources of torsion and nonmetricity with respect to the rest of the parameters present in the Riemannian case shall be straightforwardly stressed in Sec.~\ref{sec:4c}.

\subsection{The metric with cosmological constant}\label{sec:4a}
The PD space-time describes a general type D family of stationary and axisymmetric solutions of the Einstein-Maxwell equations of GR \cite{Plebanski:1976gy,Stephani:2003tm,Griffiths:2009dfa}. It mainly depends on seven physical parameters that under certain transformations and limiting procedures can be identified as a mass $m$, orbital angular momentum $a$, acceleration $\alpha$, NUT parameter $l$, cosmological constant $\Lambda$, and charges $q_{\rm e}$ and $q_{\rm m}$ of a double-aligned electromagnetic field \cite{Griffiths:2005qp}, as well as on an additional parameter $\chi$ that sets the distribution of axial singularities and avoids the occurrence of closed timelike and null geodesics \cite{Clement:2015cxa}. In four dimensions, the tracelessness property of the energy-momentum tensor of the electromagnetic field involves that the Riemannian scalar curvature of the Einstein-Maxwell theory is proportional to the cosmological constant, as it is also the case in the affinely connected metric space-time described by the gravitational action~(\ref{LagrangianIrreducible}), which fulfills the constraint $R=4\Lambda$. This special property fulfilled by the present MAG model allows us to evaluate the conditions under which the PD space-time is still valid in metric-affine geometry by the presence of both dynamical torsion and nonmetricity fields. In particular, the Expression~(\ref{LagrangianIrreducible}) extends the form of the well-known triplet type solutions covered by the Obukhov's equivalence theorem with additional quadratic curvature corrections \cite{Obukhov:1996ka,Garcia:1998jw}.

Assuming Boyer-Lindquist coordinates $(t,r,\vartheta,\varphi)$, the PD space-time with two expanding and double principal null directions can be settled from the line element
\begin{align}\label{PlebDemMetric}
\dd s^2 &=
\Omega^{-2}(r,\vartheta)\Big\{ \Phi_1(r,\vartheta)\left[\,\dd t-\left(a\sin^2\vartheta +2l(\chi-\cos\vartheta)\right)\dd\varphi \right]^2
   - \frac{\dd r^2}{\Phi_1(r,\vartheta)}  \nonumber\\
& \quad  - \,\frac{\dd\vartheta^2}{\Phi_2(r,\vartheta)}
  -\Phi_2(r,\vartheta)\sin^2\vartheta\big[ a\,\dd t -\big(r^2+a^2+l^2+2\chi a l\big)\,\dd\varphi \big]^2\Big\}\,,
\end{align}
where $\Phi_1$, $\Phi_2$ and $\Omega$ are three independent functions recovered from the general axisymmetric form~\eqref{eq:metric} when imposing
\begin{align}
\Psi_1(r,\vartheta)&=\Omega^{-2}(r,\vartheta)\Big[\Phi_1(r,\vartheta )-a^2\Phi_2(r,\vartheta)\sin^2\vartheta\Big]\,,\label{metric_mapping1}\\
\Psi_2(r,\vartheta)&=\Omega^{2}(r,\vartheta)\Phi_1(r,\vartheta)\,,\quad \Psi_3(r,\vartheta)=\frac{1}{r^2\Omega^{2}(r,\vartheta)\Phi_2(r,\vartheta)}\,,\label{metric_mapping2}\\
 \Psi_4(r,\vartheta )&=\frac{\Phi_2(r,\vartheta)}{\Psi_5(r,\vartheta)}\Big\{a\Phi_2(r,\vartheta)\left(r^2+a^2+l^2+2al\chi\right)-\Phi_1(r,\vartheta)\left[a+2l\csc^{2}\vartheta\left(\chi-\cos\vartheta\right)\right]\Big\}\,,\label{metric_mapping3}\\
 \Psi_5(r,\vartheta )&=\Big\{\Phi_{2}^{2}(r,\vartheta)\left(r^2+a^2+l^2+2al\chi\right)^2-\frac{1}{4}\Phi_1(r,\vartheta)\Phi_2(r,\vartheta)\csc^2\vartheta\left[2a\sin^{2}\vartheta+4l\left(\chi-\cos\vartheta\right)\right]^{2}\Big\}^{1/2}\,.\label{metric_mapping4}
\end{align}
In particular, by ascribing the expressions of $\Phi_1$ and $\Phi_2$ to two arbitrary functions $Q$ and $P$ depending on the radial and polar coordinates, respectively
\begin{align}
      \Phi_1(r,\vartheta)&=\frac{Q(r)}{\rho^2(r,\vartheta)}\,,\quad \Phi_2(r,\vartheta)=\frac{P(\vartheta)}{\rho^2(r,\vartheta)}\,,
\end{align}
with
\begin{equation}
    \rho^{2}(r,\vartheta)=r^2+(a\cos\vartheta+l)^2\,,
\end{equation}
then the resulting metric tensor admits a nondegenerate rank-2 conformal Killing-Yano tensor\footnote{Note that the dual tensor $*h_{\mu\nu}=(1/2)\,\varepsilon^{\lambda\rho}\,_{\mu\nu}h_{\lambda\rho}$ is also a nondegenerate rank-2 conformal Killing-Yano tensor.}
\begin{equation}\label{killing-yano_tensor}
    h_{\mu\nu}=\frac{\partial_{[\mu}b_{\nu]}}{\Omega^3\left(r,\vartheta\right)}\,,
\end{equation}
fulfilling the following equation~\cite{Frolov:2017kze}:
\begin{equation}\label{eq_K-Y}
    \nabla_{\lambda}h_{\mu\nu}=\nabla_{[\lambda}h_{\mu\nu]}+\frac{2}{3}g_{\lambda[\mu|}\nabla_{\rho}h^{\rho}\,_{|\nu]}\,,
\end{equation}
with the vector $b_\mu$ being
\begin{align}
    b_{\mu}&=\frac{1}{2}\Big\{r^2+a^2\sin^2\vartheta-2al\cos\vartheta,0,0,2\left(r^2+a^2+b^2+2\chi al\right)l\cos\vartheta-2\chi lr^2-\left(r^2+a^2+b^2+2\chi al\right)a\sin^2\vartheta\Big\}\,.
\end{align}
Hence, it is straightforward to check that the mentioned Killing-Yano tensor is closed if $\Omega(r,\vartheta)=1$, in such a way that Eq.~(\ref{eq_K-Y}) is reduced to the following expression:
\begin{equation}
    \nabla_{(\lambda}h_{\mu)\nu}=0\,,
\end{equation}
with
\begin{align}
    h_{t r}&=a\cos\vartheta+l\,, \quad h_{r \varphi}=\left(a\cos\vartheta+l\right)\left[a\sin^{2}\vartheta+2l\left(\chi-\cos\vartheta\right)\right]\,,\\
    h_{t \vartheta}&=-\,ar\sin\vartheta\,, \quad h_{\vartheta\varphi}=-\,r\sin\vartheta\left(r^2+a^2+l^2+2\chi al\right)\,.
\end{align}

The set of functions $Q$, $P$ and $\Omega$ can be considered then to contain the information of the remaining parameters and fix the PD geometry. In this regard, by considering the new parametrisation achieved by Podolsk\'y and Vr\'atn\'y with vanishing cosmological constant \cite{Podolsky:2021zwr}, the corresponding PD space-time with two expanding and double principal null directions acquires a particularly meaningful form, which allows us to introduce the cosmological constant as follows:
\begin{align}
     Q(r)&= \big(r-r_{+}\big) \big(r-r_{-}\big)
            \Big(1+\alpha\,a\,\frac{a-l}{a^2+l^2}\,r\Big)
            \Big(1-\alpha\,a\,\frac{a+l}{a^2+l^2}\,r\Big)+\frac{1}{3} \Lambda\left[\left(r^2+a^2\right)\left(r^2+3 l^2\right)+3l^2\left(r^2-l^2\right)\right]\nonumber\\
            &+\frac{1}{3}\alpha a\Lambda  \left(c_3+c_4 r+c_5 r^2+c_6 r^3+c_7 r^4\right)\,,\\
             P(\vartheta)&=\Bigl[ 1-\frac{\alpha\,a}{a^2+l^2}\, r_{+}\left(l+a\cos\vartheta\right)\Bigr]\Big[ 1-\frac{\alpha\,a}{a^2+l^2}\, r_{-}\left(l+a\cos\vartheta\right)\Big]-\frac{1}{3} a \Lambda  \cos \vartheta (a \cos \vartheta+4 l)\nonumber\\
             &+\frac{1}{3} \alpha  a \Lambda  \left(c_0+c_1 \cos \vartheta+c_2 \cos ^2\vartheta\right)\,,\\
            \Omega(r,\vartheta)&=1-\frac{\alpha a}{a^2+l^2}r\left(a\cos\vartheta+l\right)\,,
\end{align}
where the factor $r_{\pm}$ is defined as
\begin{equation}
    r_{\pm}=m\pm\sqrt{m^2+l^2-a^2-k}\,,
\end{equation}
and $\{c_{i}\}_{i=0}^{7}$ are eight constants satisfying the relations
\begin{eqnarray}
c_0&=&c_3 \biggl[3\left(\frac{\alpha al}{a^2+l^2}\right)^{2}+\frac{1}{a^2-l^2}\biggr]+\frac{a^2\left(c_4+9l^3\right)+l^2 \left(c_4-9 l^3\right)}{\left(a^2+l^2\right)^2}\,\alpha  al\,,\\
c_1&=&\frac{4c_{3}\alpha al+a^2 \left(c_4+12 l^3\right)+l^2\left(c_4-12 l^3\right)}{\left(a^2+l^2\right)^2}\,\alpha  a^2\,,\\
c_2&=&\frac{c_{3}\alpha  a+3a^2l^2-3l^4}{\left(a^2+l^2\right)^2}\,\alpha a^3\,,\\
c_5&=&c_3 \biggl[\frac{1}{a^2-l^2}-\frac{\left(a^2+3 l^2\right)}{\left(a^2+l^2\right)^2}\,\alpha^2a^2\biggr]-\frac{3l\left(a^2+3l^2\right)\left(a^2-l^2\right)+2c_4\left(a^2+l^2\right)}{\left(a^2+l^2\right)^2}\,\alpha  al\,,\\
c_6&=&-\,\frac{1}{\left(a^2-l^2\right)\left(a^2+l^2\right)^3}\Big\{2l\left(a^2+l^2\right)^2\left(5a^2l^2-a^4-4l^4\right)+2c_3\alpha al\Bigl[\left(a^2+l^2\right)^2+\alpha^2a^2\left(a^2-l^2\right)^2\Bigr]\nonumber \\
&&+\,\alpha^2a^2\left(a^2-l^2\right)^2 \Bigl[6l^3\left(a^2-l^2\right)+c_4\left(a^2+l^2\right)\Bigr]\Big\}\,, \\
c_7&=&-\,\frac{3l^2\left(a^2-l^2\right)+c_{3}\alpha a}{\left(a^2+l^2\right)^2}\,\alpha a\,.
\end{eqnarray}
Thereby, the line element presented above is aimed to encode the hypermomentum parameters of our MAG model by a constant $k$, which in the Einstein-Maxwell theory also contains the corresponding electromagnetic charges of the gauge group $U(1)$, as well as the mass into the metric function $Q(r)$. The orbital angular momentum, acceleration and NUT parameters are present in all the metric components, in such a way that the geometry becomes completely flat when expressed only in terms of $a$ and $\alpha$ (i.e. in the absence of physical sources). In addition, it also incorporates the cosmological constant into $Q(r)$ and $P(\vartheta)$, which generalises the new improved form of the PD space-time ~\cite{Podolsky:2021zwr} in the presence of a cosmological constant. As can be seen, this extension includes nontrivial contributions of the acceleration, angular momentum and NUT parameters in the metric tensor by their cross products with the cosmological constant, hence displaying the combined effects of all these parameters in the geometry, which especially affects its horizons and geodesics. The constants $c_3$ and $c_4$ act as free parameters in the metric tensor, so they can be set to any convenient value. In this sense, a full geodesic analysis of the complete PD space-time is especially desirable to clarify the geometric implications of the constants $c_3$ and $c_4$.

\subsection{Solution with dynamical torsion and nonmetricity tensors}\label{sec:4b}

The evaluation of the field equations (\ref{field_eq1}) and (\ref{field_eq2}) in the PD space-time enables a generalisation of the PD solution of GR in the framework of MAG. In this sense, as previously mentioned, the problem can be reduced to the search of the three main pieces of the spin connection (\ref{SCtotal}) in the local frame provided by the Lorentz transformation (\ref{LorentzMatrixTransformation}), which can be trivially mapped from a general stationary and axisymmetric geometry to the PD space-time by the correspondence (\ref{metric_mapping1})-(\ref{metric_mapping4}). The first part consists then of a Minkowski component (\ref{MinkowskiSP}) with vanishing field strength tensors, which identically satisfies the field equations of MAG.

The second part encodes the dynamical contribution of the Weyl vector $W_{\mu}=\left(w_{1}(r,\vartheta),w_{2}(r,\vartheta),w_{3}(r,\vartheta),w_{4}(r,\vartheta)\right)$ in a symmetric component of the spin connection, which is realised by an auxiliary vector mode of torsion fulfilling the condition $\mathring{T}_{2}^{\nu}=(3/2)\,W^{\nu}$. Accordingly, the propagating equation of nonmetricity corresponds with the symmetric component in the indices $\lambda$ and $\mu$ of Eq.~(\ref{field_eq2}), which leads to the Maxwell-like equation
\begin{equation}
    \nabla_{\mu}\tilde{R}^{\lambda}\,_{\lambda}\,^{\mu\nu}=0\,.\label{eqnon}
\end{equation}

The resolution of this equation in the PD space-time can be accomplished in a systematic way from the metric (\ref{PlebDemMetric}). First, the components $\nu=1$ and $\nu=2$ of Eq.~\eqref{eqnon} can be written as\footnote{Note that for simplicity, we omitted the dependence of the functions on $r$ and $\vartheta$.}
\begin{eqnarray}
0&=&F_0\bigg[\frac{\partial_\vartheta\Phi_1}{\Phi_1}+\frac{\partial_\vartheta\Phi_2}{\Phi_2}+\cot \vartheta-\frac{2 a \sin \vartheta (a \cos \vartheta+l)}{r^2+\left(a\cos\vartheta+l\right)^{2}}\bigg]+\partial_\vartheta F_0\,,\\
0&=&F_0\,\bigg\{\frac{\partial_r\Phi_1}{\Phi_1}+\frac{\partial_r\Phi_2}{\Phi_2}+\frac{2 r}{r^2+\left(a\cos\vartheta+l\right)^{2}}\bigg\}+\partial_r F_0\,,
\end{eqnarray}
or equivalently
\begin{equation}
    \partial_{\nu}\left\{F_{0}\Phi_1\Phi_2\bigl[r^2+\left(a\cos\vartheta+l\right)^{2}\,\bigr]\sin\vartheta\right\}=0\,,
\end{equation}
with \begin{equation}
    F_0=\partial_{\vartheta}w_{2}-\partial_{r}w_{3}\,.
\end{equation}
In order to solve these equations, we impose the function $F_0$ that relates the respective components of the Weyl vector to be independent from the metric functions $\Phi_1$ and $\Phi_2$ and therefore to be equal to zero:
\begin{equation}\label{condd}
    \partial_{\vartheta}w_{2}=\partial_{r}w_{3}\,.
\end{equation}
Note that the case with nonvanishing $F_{0}$ generally modifies the nonmetricity scalar $W_{\mu}W^{\mu}$ constructed from the Weyl vector and also induces a constant magnetic component in the azimuth direction. 

Likewise, the components $\nu=0$ and $\nu=3$ of Eq.~\eqref{eqnon} can be written in the following compact form:
\begin{eqnarray}
0&=&F_1\bigg[\frac{\partial_r\Phi_1}{\Phi_1}-\frac{\partial_r\Phi_2}{\Phi_2}-\frac{2r}{r^2+\left(a\cos\vartheta+l\right)^2}\bigg]+F_2\bigg[\cot\vartheta+\frac{\partial_\vartheta\Phi_1}{\Phi_1}-\frac{\partial_\vartheta\Phi_2}{\Phi_2}-\frac{2a\sin\vartheta\left(a\cos\vartheta+l\right)}{r^2+\left(a\cos\vartheta+l\right)^{2}}\bigg]\nonumber\\
&&+\Phi_1^2 \Phi_2 \partial_r\bigg(\frac{F_1}{\Phi_1^2 \Phi_2}\bigg)-\Phi_2^3 \partial_\vartheta\bigg(\frac{F_2}{\Phi_2^3}\bigg)+F_3 \Phi_1 \Phi_2^2\,,\label{eq38}\\
0&=&F_1\Bigg[\frac{8\Phi_{1}^3\Phi_2\csc ^2\vartheta}{a\sin^{2}\vartheta+2l\left(\chi-\cos\vartheta\right)}\Bigg]\left[\frac{\partial_r\Phi_2}{\Phi_2}-\frac{\partial_r\Phi_1}{\Phi_1}+\frac{2r}{r^2+\left(a\cos\vartheta+l\right)^2}\right]\nonumber\\
&&+\,F_2\bigg(\frac{8a\Phi_{1}^3\Phi_2\csc ^2\vartheta}{r^2+a^2+l^2+2al\chi}\bigg)\Bigg[\cot\vartheta+\frac{\partial_\vartheta\Phi_2}{\Phi_2}-\frac{\partial_\vartheta\Phi_1}{\Phi_1}+\frac{2a\sin\vartheta\left(a\cos\vartheta +l\right)}{r^2+\left(a\cos\vartheta+l\right)^{2}}\Bigg]\nonumber\\
&&-\,\Phi_1^5 \Phi_2^2 \partial_r\bigg\{\frac{F_1}{\Phi_1^2\Phi_2}\bigg[\frac{8\csc^2\vartheta}{a\sin^{2}\vartheta+2l\left(\chi-\cos\vartheta\right)}\bigg]\bigg\}+\Phi_1^3\Phi_2^4 \partial_\vartheta\bigg\{\frac{F_2}{\Phi_2^3}\bigg[\frac{8 a\csc ^2\vartheta }{r^2+a^2+l^2+2al\chi}\bigg]\bigg\}\nonumber\\
&&-\,\frac{8\left(F_3+F_4\right)\,\Phi_1^4\Phi_2^3\csc^2\vartheta}{a\sin^{2}\vartheta+2l\left(\chi-\cos\vartheta\right)}\,,\label{eq39}
\end{eqnarray}
where the functions $\{F_{i}\}_{i=1}^{4}$ are defined as
\begin{eqnarray}
F_1(r,\vartheta)&=&-\,\Phi_1^2\Phi_2\Big\{\left[a\sin^{2}\vartheta+2l\left(\chi-\cos\vartheta\right)\right]\partial_r w_1+\partial_r w_4\Big\}\left[a\sin^{2}\vartheta+2l\left(\chi-\cos\vartheta\right)\right]\,,\\
F_2(r,\vartheta)&=&-\,\Phi_2^3\Big[ \left(r^2+a^2+l^2+2al\chi\right)\partial_\vartheta w_1+a\partial_\vartheta w_4\Big]\left(r^2+a^2+l^2+2al\chi\right)\sin^2\vartheta\,,\\
F_3(r,\vartheta)&=&\left(\frac{a \partial_{rr}w_4}{r^2+a^2+l^2+2al\chi}+\partial_{rr}w_1\right)\left(r^2+a^2+l^2+2al\chi\right)^2\sin^2\vartheta+g_1\partial_\vartheta w_1+g_2\partial_\vartheta w_4\nonumber\\
&&-\,\Big\{\left[a\sin^{2}\vartheta+2l\left(\chi-\cos\vartheta\right)\right]\partial_{\vartheta\vartheta}w_{1}+\partial_{\vartheta\vartheta}w_4\Big\}\left[a\sin^{2}\vartheta+2l\left(\chi-\cos\vartheta\right)\right]\nonumber\\
&&+\,\frac{2r\sin^2\vartheta\left(r^2+a^2+l^2+2al\chi\right) \left(a^2 \cos (2 \vartheta )+4 al \cos \vartheta -2 al \chi +l^2+r^2\right)}{r^2+\left(a\cos\vartheta+l\right)^2}\,\partial_r w_1\nonumber\\
&&-\,\frac{2a^{2}r\sin^2\vartheta\left[a\sin^{2}\vartheta+2l\left(\chi-\cos\vartheta\right)\right]}{r^2+\left(a\cos\vartheta+l\right)^2}\,\partial_r w_4\,,\label{F3}\\
F_4(r,\vartheta)&=&-\,\Big\{\sin\vartheta\left[r^2+\left(a\cos\vartheta+l\right)^{2}\right]\Big[\left(r^2+a^2+l^2+2al\chi\right)\partial_{rr}w_1 +a\partial_{rr}w_4+2r\partial_{r}w_1\Big]\nonumber\\
&&-\,2\left(a\cos\vartheta+l\right)\left[a\sin^{2}\vartheta+2l\left(\chi-\cos\vartheta\right)\right]\partial_\vartheta w_1-2(a \cos \vartheta+l) \partial_\vartheta w_4 \Big\}\sin\vartheta\,,
\end{eqnarray}
with
\begin{eqnarray}
g_1(r,\vartheta)&=&2\left[a\sin^{2}\vartheta+2l\left(\chi-\cos\vartheta\right)\right]\Big[a^3 \cos (5 \vartheta )+3 a \cos (3 \vartheta ) \left(3 a^2+8 a l \chi +4 \left(l^2+r^2\right)\right)\nonumber\\
&&+8 l \chi  \cos \vartheta  \left(a^2+4 \left(l^2+r^2\right)\right)-2 a \cos \vartheta  \left(5 a^2+38 l^2+6 r^2\right)-12 l \left(3 a^2+4 \left(l^2+r^2\right)\right)+4 a^2 l \cos (4 \vartheta )\nonumber\\
&&+16 l \cos(2\vartheta) \left(4 a l \chi +l^2+r^2\right)\Big]\times\Big\{32\sin\vartheta\Bigl[r^2+\left(a\cos\vartheta+l\right)^2\Bigr]\Big\}^{-1}\,,\\
g_2(r,\vartheta)&=&-\,\Big[a \left(a^2 \cos (5 \vartheta )-\cos (3 \vartheta ) \left(7 a^2+24 al \chi -12 l^2+4 r^2\right)+8 a l \cos (4 \vartheta )-64 l^2 \chi  \cos (2 \vartheta )\right)+8 l \left(3 a^2+4 \left(l^2+r^2\right)\right)\nonumber\\
&&+\cos\vartheta  \left(6 a^3-8 l \chi  \left(a^2+4 \left(l^2+r^2\right)\right)+4 a \left(13 l^2+r^2\right)\right)\Big]\Big\{16\sin\vartheta\Bigl[r^2+\left(a\cos\vartheta+l\right)^2\Bigr]\Big\}^{-1}\,.
\end{eqnarray}
In order to solve the system~\eqref{eq38}-\eqref{eq39} for any arbitrary value of the functions $\Phi_1$ and $\Phi_2$, we must impose $F_{i}=0$. For the specific case $F_1=F_2=0$, we find
\begin{eqnarray}
w_1(r,\vartheta)&=&w_{1b}(r)-\frac{a w_4(r,\vartheta )}{r^2+a^2+l^2+2al\chi}\,,\\
w_4(r,\vartheta)&=&\frac{\left(r^2+a^2+l^2+2\chi al\right)\left\{w_{4b}(\vartheta)-2w_{1b}(r)\left[a\sin^{2}\vartheta+2l\left(\chi-\cos\vartheta\right)\right]\right\}}{2\Bigl[r^2+\left(a\cos\vartheta+l\right)^2\Bigr]}\,,
\end{eqnarray}
where $w_{1b}(r)$ and $w_{4b}(\vartheta)$ are integration functions. The last equations $F_3=F_4=0$ restrict the form of the integration functions to be associated with two additional integration constants $\kappa_{\rm d,e}$ and $\kappa_{\rm d,m}$ that represent the electric and magnetic dilation charges
\begin{eqnarray}
w_{1b}(r)&=&\frac{\kappa_{\rm d,e}r-\kappa_{\rm d,m}\left(a\gamma+l\right)}{r^2+a^2+l^2+2\chi al}\,,\quad w_{4b}(\vartheta)=2\kappa_{\rm d,m}\left(\cos\vartheta-\gamma\right)\,,
\end{eqnarray}
whereas the constant $\gamma$ ensures the absence of axial singularities in the Weyl vector if $\gamma = \pm 1$. As is shown, the resolution of the field equations for the nonmetricity sector does not univocally determine the four components of the Weyl vector in the presence of a stationary and axially symmetric space-time, since the functions $w_2$ and $w_3$ are simply constrained by the Eq.~(\ref{condd}). By attending to the regularity of the nonmetricity invariants we can fix their values with a regular scalar $W_{\mu}W^{\mu}$, which allows us to obtain the final solution
\begin{align}
    w_{1}(r,\vartheta)&=\frac{\kappa_{\rm d,e}r-\kappa_{\rm d,m}(a\cos\vartheta+l)}{r^2+\left(a\cos\vartheta+l\right)^{2}}\,, \\
    w_2(r,\vartheta)&=-\,\frac{\kappa_{\rm d,e}r-\kappa_{\rm d,m}\left(a\gamma+l\right)}{ Q(r)}\,,\\
    w_3(r,\vartheta)&=-\,\kappa_{\rm d,m}\sqrt{K(\vartheta)-\left(\frac{\cot \vartheta-\gamma\csc\vartheta}{P(\vartheta)}\right)^2}\,,\\
    w_{4}(r,\vartheta)&=\kappa_{\rm d,m}\left[\frac{\left(r^2+a^2-l^2\right)\cos\vartheta+al\sin^{2}\vartheta+2\chi l\left(a\cos\vartheta+l\right)}{r^2+\left(a\cos\vartheta+l\right)^2}-\gamma\right]-\frac{\kappa_{\rm d,e}r\left[a\sin^{2}\vartheta+2l\left(\chi-\cos\vartheta\right)\right]}{r^2+\left(a\cos\vartheta+l\right)^2}\,,
\end{align}
where $K(\vartheta)$ is a new integration function fulfilling the condition $K(\vartheta) \geq \left(\cot \vartheta-\gamma\csc\vartheta\right)^2/P^2(\vartheta)$. Thereby, the quadratic scalar invariant constructed from the Weyl vector reads
\begin{equation}
    W_{\mu}W^{\mu}=-\,\frac{\kappa_{\rm d,m}^{2}K(\vartheta)P(\vartheta)\Omega^{2}(r,\vartheta)}{r^2+\left(a\cos\vartheta+b\right)^2}\,.
\end{equation}

In order to characterise this type of monopole configuration, we calculate the form of the electric $E_{\mu}=\tilde{R}^{\lambda}\,_{\lambda\mu\nu}\,u{^\nu}$ and magnetic components $B_{\mu}=(1/2)\,\varepsilon_{\mu\nu\lambda\rho}\tilde{R}_{\sigma}\,^{\sigma\lambda\rho}\,u{^\nu}$, as measured by an observer moving with four-velocity $u{^\nu}$. In particular, we find the following electric and magnetic projections of the homothetic field strength tensor in the rest frame of reference:
\begin{align}\label{ElectricMagneticComponents}
    E_{r}&=\frac{2\kappa_{\rm d,e}\left[r^2-\left(a\cos\vartheta+l\right)^2\right]-4\kappa_{\rm d,m}r\left(a\cos\vartheta+l\right)}{\Bigl[r^2+(a\cos\vartheta+l)^2\Bigr]^2}\,,\\
    E_{\vartheta}&=-\,\frac{4\kappa_{\rm d,e}r\left(a\cos\vartheta+l\right)+2\kappa_{\rm d,m}\left[r^2-\left(a\cos\vartheta+l\right)^2\right]}{\Bigl[r^2+(a \cos\vartheta+l)^{2}\Bigr]^2}\,a\sin\vartheta\,,\\
    B_{r}&=-\,\frac{2\kappa_{\rm d,m}\left[r^2-\left(a\cos\vartheta+l\right)^2\right]+4\kappa_{\rm d,e}r\left(a\cos\vartheta+l\right)}{\Bigl[r^2+(a \cos \vartheta+l)^2\Bigr]^2}\,,\\
    B_{\vartheta}&=\frac{4\kappa_{\rm d,m}r\left(a\cos\vartheta+l\right)-2\kappa_{\rm d,e}\left[r^2-\left(a\cos\vartheta+l\right)^2\right]}{\Bigl[r^2+(a\cos\vartheta+l)^2\Bigr]^2}\,a\sin\vartheta\,.
\end{align}

Finally, the antisymmetric components of the Expression~(\ref{field_eq2}) provide the field equation for the torsion tensor described by the third part of the spin connection in the PD space-time. In this sense, it is straightforward to note that this equation can be expressed as the divergences of the antisymmetrized part of the curvature tensor and of the antisymmetric component of the Ricci (and equivalently of the co-Ricci tensor, provided the Maxwell-like equation~(\ref{eqnon}) for the nonmetricity tensor) plus a more complex differential expression involving nonlinear corrections in the mentioned parts of curvature and in the distortion tensor. The obtainment of an exact solution in the dynamical torsion sector requires then the fulfillment of the full differential expression with such nonlinear contributions, which account for the interaction between the physical sources of the PD space-time and torsion. The simplest case is then described in the decoupling limit, where certain conditions satisfied by the parameters of the solution allow this interaction to be neglected and the antisymmetric component of the Expression~(\ref{field_eq2}) is reduced to the Maxwell-like equations
\begin{eqnarray}
\nabla_{\lambda}\tilde{R}^{\lambda}\,_{[\rho \mu \nu]}&=&0\label{Maxwell1}\,,\\
\nabla_{\mu}\tilde{R}^{[\mu\nu]}&=&0\label{Maxwell2}\,.
\end{eqnarray}

These properties match with a Coulomb-like torsion tensor depending on a spin charge $\kappa_{s}$ in the limit $|x_{i}\kappa_{s}| \ll 1$, with $x_{i}=\left(a,l,\alpha\right)$:
\begin{align}
    \bar{T}^{\vartheta}\,_{\varphi t}=-\,\bar{T}^{\varphi}\,_{\vartheta t}\sin^{2}\vartheta=-\,\bar{T}^{\vartheta}\,_{\varphi r}\frac{Q(r)}{\rho^2(r,\vartheta)}=\bar{T}^{\varphi}\,_{\vartheta r}\frac{Q(r)}{\rho^2(r,\vartheta)}\sin^{2}\vartheta=\frac{\kappa_{s}\sin\vartheta}{r}+\mathcal{O}(x_{i}\kappa_{s})\,,
\end{align}
which gives rise to the field strength tensors
\begin{eqnarray}
\tilde{R}^{\lambda}\,_{[\mu\nu\rho]}&=&\frac{1}{4}\tilde{R}^{\sigma}\,_{\sigma[\mu\nu}\delta_{\rho]}\,^{\lambda}+\bar{R}^{\lambda}\,_{[\mu\nu\rho]}\,,\\
\tilde{R}_{[\mu\nu]}&=&\frac{1}{4}\tilde{R}^{\sigma}\,_{\sigma\mu\nu}+\bar{R}_{[\mu\nu]}\,,\\
\bar{R}^{\lambda}\,_{[\mu \nu \rho]}&=&\frac{1}{6}\varepsilon^{\lambda}\,_{\sigma[\rho\nu}\nabla_{\mu]}\bar{S}^{\sigma}+\nabla_{[\mu}\bar{t}^{\lambda}\,_{\rho\nu]}+\frac{1}{4}\varepsilon^{\lambda}\,_{\omega\sigma[\rho}\mathring{t}_{1}^{\sigma}\,_{\mu\nu]}\bar{S}^{\omega}-\frac{1}{18}\varepsilon_{\sigma\mu\nu\rho}\mathring{T}_{1}^{\lambda}\bar{S}^{\sigma}\,,\\
\bar{R}_{[\mu\nu]}&=&\frac{1}{12}\varepsilon^{\lambda}\,_{\sigma\mu\nu}\nabla_{\lambda}\bar{S}^{\sigma}+\frac{1}{2}\nabla_{\lambda}\bar{t}^{\lambda}\,_{\mu\nu}\,.
\end{eqnarray}

Thereby, besides the Riemannian structure described by the metric tensor, the final configuration consists of two dynamical axial and tensor modes of torsion, as well as of a dynamical Weyl vector related to the trace part of nonmetricity. The remaining field equation~(\ref{field_eq1}) associates the constant $k$ present in the metric function $Q(r)$ with the hypermomentum charges
\begin{equation}
    k=d_{1}\kappa_{\rm s}^{2}-4e_{1}\left(\kappa_{\rm  d,e}^{2}+\kappa_{\rm d,m}^{2}\right)\,.
\end{equation}
It is straightforward to check then that the contribution of these charges in the metric tensor is analogous to the one provided by the electromagnetic charges $q_{\rm e}$ and $q_{\rm m}$ of the gauge group $U(1)$~\cite{Griffiths:2009dfa}, in such a way that the complete solution with the full list of parameters simply extends the constant $k \rightarrow k = q_{\rm e}^{2}+q_{\rm m}^{2}+d_{1}\kappa_{\rm s}^{2}-4e_{1}\bigl(\kappa_{\rm d,e}^{2}+\kappa_{\rm d,m}^{2}\bigr)$.

Overall, the tuple composed by the full anholonomic connection:
\begin{equation}\label{connection_sol}
    \hat{\omega}=\sum_{i=1}^{3}\hat{\omega}_{i}\,,
\end{equation}
with
\begin{eqnarray}
    \hat{\omega}_{1}&=&\frac{1}{2}\bigl(J_{\hat{1}\hat{2}}-J_{\hat{0}\hat{2}}\bigr)\,\dd\vartheta +\frac{1}{2}\sin\vartheta \left[\bigl(J_{\hat{1}\hat{3}}-J_{\hat{0}\hat{3}}\bigr)+2\cot\vartheta J_{\hat{2}\hat{3}}\right] \dd\varphi\,,\label{part1}\\
    \hat{\omega}_{2}&=&\frac{1}{2}
\biggl\{
-\,\frac{\kappa_{\rm d,e}r-\kappa_{\rm d,m}(a\cos\vartheta+l)}{r^2+\left(a\cos\vartheta+l\right)^{2}}\,\dd t+\frac{\kappa_{\rm d,e}r-\kappa_{\rm d,m}\left(\pm a+l\right)}{Q(r)}\,\dd r
\biggr.
\nonumber\\
\biggl.
&&+\,\kappa_{\rm d,m}\sqrt{K(\vartheta)-\left(\frac{\cot \vartheta-\gamma\csc\vartheta}{P(\vartheta)}\right)^2}\,\dd\vartheta+\frac{\kappa_{\rm d,e}r\left[a\sin^{2}\vartheta+2l\left(\chi-\cos\vartheta\right)\right]}{r^2+\left(a\cos\vartheta+l\right)^2}\,\dd \varphi
\biggr.
\nonumber\\
\biggl.
&&+\,\kappa_{\rm d,m}\frac{\pm\bigl[r^2+\left(a\cos\vartheta+l\right)^2\bigr]-\left[\left(r^2+a^2-l^2\right)\cos\vartheta+al\sin^{2}\vartheta+2\chi l\left(a\cos\vartheta+l\right)\right]}{r^2+\left(a\cos\vartheta+l\right)^2}\,\dd\varphi
\biggr\}D\,,\label{part2}\\
    \hat{\omega}_{3}&=&-\,\frac{\kappa_{s}}{r}\left[\dd t-\frac{\rho^2(r,\vartheta)}{Q(r)}\,\dd r\right]J_{\hat{2}\hat{3}}\,,\label{part3}
\end{eqnarray}
and the rotated local frame $\vartheta^a{}=\Lambda^a{}_b \, e^b$ obtained by the application of the Lorentz transformation~(\ref{LorentzMatrixTransformation}) over the vierbein
\begin{eqnarray}
e^0&=&\frac{\sqrt{Q(r)+P(\vartheta)\left(K_1^{2}(r,\vartheta)\Omega^{4}(r,\vartheta)-a^2\sin^2\vartheta\right)}}{\rho (r,\vartheta)\Omega(r,\vartheta)}\,\dd t+\frac{K_2(r,\vartheta)\sin\vartheta}{r\rho^{2}(r,\vartheta)\Omega^{2}(r,\vartheta)}\Big\{P(\vartheta)\left(r^2+a^2+l^2+2\chi al\right)^2\nonumber\\
&&-\,Q(r)\left[a\sin^{2}\vartheta+2l\left(\chi-\cos\vartheta\right)\right]^2\csc^{2}\vartheta\Big\}\,\dd \varphi\,,\\
e^1&=&\frac{\rho(r,\vartheta)}{\Omega(r,\vartheta)\sqrt{Q(r)}}\,\dd r\,,\\
e^2&=&\frac{\rho(r,\vartheta)}{\Omega(r,\vartheta)\sqrt{P(\vartheta)}}\,\dd \vartheta\,,\\
e^3&=&\frac{K_1(r,\vartheta)\Omega(r,\vartheta)\sqrt{P(\vartheta)}}{\rho(r,\vartheta)}\,\dd t+\frac{\sin\vartheta\sqrt{K_2^{2}(r,\vartheta)\rho^{2}(r,\vartheta)+r^{2}P(\vartheta)\Omega^{2}(r,\vartheta)}}{r\rho^{3}(r,\vartheta)\Omega^{2}(r,\vartheta)}\Big\{P(\vartheta)\left(r^2+a^2+l^2+2\chi al\right)^2\nonumber\\
&&-\,Q(r)\left[a\sin^{2}\vartheta+2l\left(\chi-\cos\vartheta\right)\right]^{2}\csc^2\vartheta\Big\}\,\dd \varphi\,,
\end{eqnarray}
with
\begin{eqnarray}
    K_1(r,\vartheta)&=&\frac{K_2(r,\vartheta)\rho(r,\vartheta)\sqrt{Q(r)} \left[r^2+(a\cos\vartheta+l)^2\right]\sin\vartheta}{r\Omega^{3}(r,\vartheta)\sqrt{P^{2}(\vartheta)\left(r^2+a^2+l^2+2\chi al\right)^{2}\sin^2\vartheta-P(\vartheta)Q(r)\left[a\sin^{2}\vartheta+2l\left(\chi-\cos\vartheta\right)\right]^{2}}}\nonumber\\
    &&-\frac{\sin\vartheta  \sqrt{K_2^{2}(r,\vartheta)\rho^{2}(r,\vartheta)+r^2P(\vartheta)\Omega^{2}(r,\vartheta)}}{r P(\vartheta)\Omega^{3}(r,\vartheta)\sqrt{P(\vartheta)\big(r^2+a^2+l^2+2\chi al\bigr)^2-Q(r)\left[a\sin^{2}\vartheta+2l\left(\chi-\cos\vartheta\right)\right]^{2}\csc^2\vartheta}}\times\nonumber\\
    &&\times\Bigl\{aP(\vartheta)\bigl(r^2+a^2+l^2+2\chi al\bigr)-Q(r)\left[a\sin^{2}\vartheta+2l\left(\chi-\cos\vartheta\right)\right]\csc^2\vartheta\Big\}\,,
\end{eqnarray}
provides the torsion and nonmetricity tensors which solve the field equations~(\ref{field_eq1})-(\ref{field_eq2}) in the decoupling limit \;\;\,$|x_{i}\kappa_{\rm s}| \ll 1$, in virtue of the correspondence with the affine connection
\begin{equation}
\tilde{\Gamma}^{\lambda}\,_{\rho \mu}=\vartheta_{a}\,^{\lambda}\vartheta^{b}\,_{\rho}\,\hat{\omega}^{a}\,_{b\mu}+\vartheta_{a}\,^{\lambda}\,\partial_{\mu}\,\vartheta^{a}\,_{\rho}\,.
\end{equation}
Note that the form of this correspondence holds for any local frame, but the tuple~$\left(\vartheta^{a}\,_{\mu},\hat{\omega}_{1}^{a}\,_{b\mu}\right)$ obtained by our method acquires a strongly simplified and compact form that allows us to address the study of axial symmetry in MAG.

\subsection{Geometric effect of the spin and dilation charges}\label{sec:4c}

The general decomposition of the anholonomic connection~(\ref{connection_sol}) associates the spin and dilation charges with its antisymmetric and symmetric dynamical components, respectively. This correspondence involves a direct geometric effect on the motion of spinning and dilational test bodies characterised by a general coupling to the torsion and nonmetricity tensors~\cite{Puetzfeld:2007hr,Bahamonde:2021akc}:
\begin{equation}\label{motion}
\dot{p}^{\mu}+\Gamma^{\mu}\,_{\lambda \rho}\,p^{\lambda}u^{\rho}+N_{[\lambda \rho]}\,^{\mu}p^{\rho}u^{\lambda}+\tilde{R}_{\lambda \rho \sigma}\,^{\mu}\bigtriangleup^{\rho\lambda}u^{\sigma}=0\,,
\end{equation}
where $\bigtriangleup^{\rho\lambda}$ represents the hypermomentum tensor containing the spin and dilation currents of the test body. Therefore, such an effect is realised by Lorentz-like forces that define a set of nongeodesic trajectories for test bodies with microstructure.

For ordinary matter uncoupled to torsion and nonmetricity, the antisymmetric part of the canonical energy-momentum tensor and the hypermomentum tensor present in Eq.~(\ref{motion}) vanish, but the geometric effect of the spin and dilation charges is also encoded by the constant $k$ in the resulting geodesic motion via the Levi-Civita connection. In any case, one interesting feature arising from the extension to post-Riemannian geometry is that the sign of the constant $k$ can differ from the corresponding Riemannian solution for a sufficiently large negative contribution involving the hypermomentum and electromagnetic charges, since the axial mode of torsion is characterised by an opposite parity and kinetic term in comparison with the rest of modes of the nonmetricity and electromagnetic fields. In addition, it is worthwhile to stress that even though such geodesic paths are preserved under projective transformations of the affine connection~\cite{zbMATH02588267,haantjes1937projective}, the gravitational action~\eqref{LagrangianIrreducible} is not invariant under these transformations, which means that in the present case neither the torsion tensor nor the nonmetricity tensor appear as purely projective modes but they represent totally independent dynamical properties of the space-time~\cite{BeltranJimenez:2017doy,Afonso:2017bxr} (see~\cite{BeltranJimenez:2019acz,BeltranJimenez:2020sqf} for a thorough analysis on the role of projective symmetry in the stability of metric-affine models).

On the other hand, the roots of the metric function $Q(r)$ determine the black hole, cosmological and acceleration horizons. In this sense, an analytical expression of the black hole and acceleration horizons can be straightforwardly derived for the case $\Lambda=0$:
\begin{eqnarray}
    r^{(\rm bh)}_{\pm}&=&m \pm \sqrt{m^{2}+l^{2}-a^{2}-q_{\rm e}^{2}-q_{\rm m}^{2}-d_{1}\kappa^{2}_{\rm s}+4e_{1}\kappa^{2}_{\rm d}}\,,\\
    r^{(\rm ac)}_{\pm}&=&\pm\,\alpha^{-1}(a^2+l^2)/(a^2 \pm al)\,,
\end{eqnarray}
which for real positive values of $r^{(\rm bh)}_{\pm}$ and $r^{(\rm ac)}_{\pm}$ provides two black hole horizons and two acceleration horizons. Likewise, the equation
\begin{equation}
    Q(r_{e})-a^{2}P(\vartheta)\sin^{2}\vartheta=0\,,
\end{equation}
describes the hypersurfaces $r_{e}(\vartheta)$ limiting the corresponding horizons and the static limit, or ergosurfaces. It is clear then that the presence of the electromagnetic and hypermomentum charges modifies the position of the black hole horizons and the ergosurfaces but, as previously stressed, for sufficiently high values of the spin charge and also of the NUT parameter fulfilling the condition
\begin{equation}\label{cond_1}
    l^{2}-a^{2}-q_{\rm e}^{2}-q_{\rm m}^{2}-d_{1}\kappa^{2}_{\rm s}+4e_{1}\kappa^{2}_{\rm d} > 0\,,
\end{equation}
these parameters can give rise to a negative quadratic contribution in $Q(r)$ and cancel out the inner black hole horizon. In addition, the inner and outer acceleration horizons exist for nonvanishing values of the acceleration if the following expressions are satisfied:
\begin{eqnarray}
     &\textrm{for} \; \alpha > 0\,:& r^{(\rm ac)}_{+} > 0 \quad \textrm{if}\quad \frac{l}{a} > -\,1 \,,\label{ac_cond1}\\
     &&r^{(\rm ac)}_{-} > 0\,\quad \textrm{if}\quad \frac{l}{a} > 1\,,\label{ac_cond3}\\
     &\textrm{for} \; \alpha < 0\,:& r^{(\rm ac)}_{+} > 0\,\quad \textrm{if} \quad \frac{l}{a} < -\,1\,,\label{ac_cond3}\\
     &&r^{(\rm ac)}_{-} > 0 \quad \textrm{if} \quad \frac{l}{a} < 1\,.\label{ac_cond4}
\end{eqnarray}

Furthermore, a direct computation of the Riemannian Kretschmann scalar splits the corresponding contributions of the quadratic Weyl and Ricci invariants to the geometry
\begin{equation}
 R_{\lambda\rho\mu\nu}R^{\lambda\rho\mu\nu}=48 \,\frac{\Omega^6(r,\vartheta)}{\rho^{12}(r,\vartheta)}\Bigl[K_+(r,\vartheta)K_-(r,\vartheta) +\frac{1}{6}k^2\Omega^2(r,\vartheta)\rho^4(r,\vartheta)\Bigr]\,,
\end{equation}
where we have defined
\begin{eqnarray}
K_\pm (r,\vartheta) &=& m \Big[ F_\pm(r,\vartheta) \pm \alpha\,a\,\frac{a^2-l^2}{a^2+l^2}\, F_\mp(r,\vartheta) \Big]
 \mp l \Big[ F_\mp(r,\vartheta) \mp \alpha\,a\,\frac{a^2-l^2+k}{a^2+l^2}\, F_\pm (r,\vartheta) \Big]\nonumber\\
 &&-\,k\Big[1+\frac{\alpha\,a}{a^2+l^2}\,r \left(a\cos\vartheta+l\right)\Big] \, T_\pm (r,\vartheta)\,, \\
F_\pm (r,\vartheta) &=& \big[r\mp \left(a\cos\vartheta+l\right)\big] \big[r^2+\left(a\cos\vartheta+l\right)^2\pm 4r\left(a\cos\vartheta+l\right)\big]\,, \\
T_\pm (r,\vartheta) &=& \big[r^2-\left(a\cos\vartheta+l\right)^2\pm 2r\left(a\cos\vartheta+l\right)\big]\,.
\end{eqnarray}
This quantity provides an additional condition for the parameters to avoid the presence of curvature singularities at the roots of $\rho(r,\vartheta)$, namely
\begin{equation}\label{cond_2}
    |l| > |a| \geq 0\,.
\end{equation}

Therefore, the fulfillment of the Expressions~(\ref{cond_1}) and~(\ref{cond_2}) in the absence of a cosmological constant gives rise to a regular PD solution endowed with a unique outer black hole horizon and two acceleration horizons, provided the conditions~(\ref{ac_cond1})-(\ref{ac_cond4}).


\section{Conclusions}\label{sec:conclusions}
In the present work, we have extended the correspondence between rotating Kerr-Newman space-times with the torsion and nonmetricity tensors to the PD geometry \cite{Bahamonde:2021qjk}, which constitutes the most general type D configuration with two expanding and double principal null directions of the Einstein-Maxwell model of GR. For this task, we have considered the new improved form of the PD metric where all its key functions are fully explicit and factorised \cite{Podolsky:2021zwr}. As pointed out in \cite{Podolsky:2021zwr}, the new parametrisation allows a more clear and natural evaluation of the physical properties of the solution, which also simplifies its extension towards metric-affine geometry. The application of the invariance conditions related to stationarity and axial symmetry over the torsion and nonmetricity tensors leads to twenty four and four components in Weyl-Cartan geometry, respectively, which means that additional assumptions are required to address the problem. Specifically, we split the spin connection into three different parts that encode their corresponding dynamical contributions to the field strength tensors and solve the field equations of the model in a systematic way.

The resulting generalised PD solution describes a uniformly accelerated rotating black hole with a NUT parameter, cosmological constant, electromagnetic charges, and spin and dilation charges related to torsion and nonmetricity. Therefore, the contribution of the torsion and nonmetricity tensors to the geometry are perfectly distinguishable from the rest of physical degrees of freedom and the solution reduces to the standard PD solution of the Einstein-Maxwell equations in the presence of a cosmological constant when their dynamical role is switched off. The former is subject to a decoupling limit $|x_{i}\kappa_{\rm s}| \ll 1$ with the three parameters $x_{i}=(a,l,\alpha)$ responsible of axial symmetry, beyond which the nonlinear interaction between the spin angular momentum and these parameters is expected to be enough strong to correct the PD geometry. In this sense, the presented parts~(\ref{part1})-(\ref{part3}) of the spin connection in the rotated frame displayed by the Lorentz transformation~(\ref{LorentzMatrixTransformation}) constitute the starting point for the extension of the solution beyond the decoupling limit.

On the other hand, the role of the spin and dilation charges in the PD geometry is similar to the one displayed by the electromagnetic charges. An analytical expression for the black hole and acceleration horizons can be obtained by switching off the cosmological constant, which clearly shows that for negative values of $d_{1}\kappa^{2}_{\rm s}-4e_{1}\kappa^{2}_{\rm d}$ their contribution in the black hole horizons $r^{(\rm bh)}_{\pm}=m \pm \sqrt{m^{2}+l^{2}-a^{2}-q_{\rm e}^{2}-q_{\rm m}^{2}-d_{1}\kappa^{2}_{\rm s}+4e_{1}\kappa^{2}_{\rm d}}$ is the same as the one provided by the NUT parameter rather than the one of the electromagnetic charges. In particular, it is worthwhile to stress that such a negative branch can cancel out the inner Cauchy horizon if $l^{2}-a^{2}-q_{\rm e}^{2}-q_{\rm m}^{2}-d_{1}\kappa^{2}_{\rm s}+4e_{1}\kappa^{2}_{\rm d} > 0$, which is characterised by the breakdown of predictability as well as by the presence of mass-inflation and kink instabilities \cite{Poisson:1990eh,Maeda:2005yd}, leading to a PD space-time endowed with a unique event horizon located at the positive root $r^{(\rm bh)}_{+}$. Conversely, none of the spin and dilation charges alters the acceleration horizons $r^{(\rm ac)}_{\pm}=\pm\,\alpha^{-1}(a^2+l^2)/(a^2 \pm al)$, which are fully determined by the orbital angular momentum, acceleration and NUT parameters, preserving the same structure as in the Riemannian case. From a phenomenological point of view, the latter plays the role of a gravitomagnetic monopole moment, in virtue of its physical effect on the geodesic motion of test bodies~\cite{Zimmerman:1989kv,LyndenBell:1996xj}. In any case, only the hypermomentum and electromagnetic charges of the solution can produce nongeodesic effects in the motion of charged test bodies with microstructure~\cite{Puetzfeld:2007hr}.

The avoidance of curvature singularities is achieved if $|l| > |a| \geq 0$, which means that the requirement of regularity and the absence of pathological inner horizons can be mainly driven by the NUT parameter and the spin charge, provided the conditions above. In this regard, the search of the full solution beyond the decoupling limit is especially relevant to evaluate how the corresponding higher order spin moments can modify such conditions. Research following these lines will be addressed in future works.

\bigskip
\bigskip
\noindent
\section*{Acknowledgements}
S.B. is supported by JSPS Postdoctoral Fellowships for Research in Japan and KAKENHI Grant-in-Aid for
Scientific Research No. JP21F21789. S.B. and L.J. acknowledge the Estonian Research Council grants PRG356 ``Gauge Gravity"  and the European Regional Development Fund through the Center of Excellence TK133 ``The Dark Side of the Universe". J.G.V. is supported by the European Regional Development Fund and the programme Mobilitas Pluss (Grant No. MOBJD541). J.G.V. is also grateful for the warm hospitality and support of the Nordic Institute for Theoretical Physics during the NORDITA Winter School 2022, where this work was completed and written.
\newpage

\bibliographystyle{utphys}
\bibliography{references}

\end{document}